\documentclass[aps,pra,showpacs,showkeys,twocolumn,%
groupedaddress,amsmath,amssymb,floatfix]{revtex4-1}
\usepackage[russian]{babel}
\usepackage{graphicx}
\usepackage{bm}
\usepackage{dcolumn}
\usepackage{epstopdf}
\usepackage{soul,color}

\begin{document}

\title{Absolute cross sections of ultra-soft x-ray radiation at electron scattering by atoms and soft-photons approximation}

\author{A. S. Kornev}
\affiliation{Voronezh State University, 394018 Voronezh, Russia}

\author{B. A. Zon}
\affiliation{Voronezh State University, 394018 Voronezh, Russia}

\author{M. Ya. Amusia}
\affiliation{The Hebrew University, Jerusalem, Israel\\ A F Ioffe Physical-Technical Institute, St. Petersburg, Russia}

\date{\today}

\begin{abstract}
We calculated the cross sections of ultra-soft x-ray bremsstrahlung in electron scattering by Ar, Kr and Xe. The results are consistent with the absolute values of the differential cross sections measured by Gnatchenko \textit{et al} [Phys. Rev. A \textbf{80}, 022707 (2009)] for scattering electrons with an energy of 600 eV on these atoms.
\end{abstract}



\maketitle

\section{Introduction}

The first experimental measurements of the relative probability of the bremsstrahlung (BS) in the long-wave region
of the spectrum in the scattering of electrons in thin metal targets were performed in the Ref. \cite{Peterson}.
The experimental data were well described by Sommerfeld's theory \cite{Som}, which is actually the theory of BS at
the Coulomb center, taking into account the exponential shielding of the nucleus by atomic electrons \cite{AB}.
The corresponding computations were carried out in Ref.~\cite{Kirk}. 
Absolute measurements of BS intensity for atomic targets (Ne, Ar, Kr, Xe) were performed for the first time, as far as we know, in Ref.~\cite{Quarles-03}. Incident electron energies in~\cite{Quarles-03} were 28 and 50 keV. The experimental results proved to be in qualitative agreement with the theory that takes into account not only the traditional BS mechanism~\cite{Som} (ordinary BS), but also the polarization BS, in which a photon is emitted by an atomic target \cite{Tsyt, Astapenko, KS-book}. However, experimental and theoretical results quantitatively differ quite significantly, and this difference increases with the BS photon energy decrease.

The recent paper \cite{Brazil} presents results of highly accurate, $\simeq 5.5$\%,
absolute cross sections of BS in the process of scattering of electrons with the energy of 20..100~keV upon thin targets C, Al, Te, Ta, and Au. This article also provides references to previous absolute measurements for thin solid targets. The results obtained in \cite{Brazil} are consistent with the theoretical data presented in the Refs.~\cite{Pratt77,Pratt81,Seltzer}. However, the authors of
\cite{Brazil} claimed that the accuracy of the existing BS theory is insufficient to compare it with high-precision experiments.
The measurements in Ref.~\cite{Brazil} did not observed polarization BS. However, this fact is not surprising, since the atoms of the used targets do not have large polarizabilities in the regions of the BS photon frequencies under consideration. It is noteworthy that in \cite{Brazil}, as in \cite{Quarles-03}, the agreement between theory and experiment worsens as the photon energy decreases. This statement is considered by the authors of Ref.~\cite{Brazil} as one of the most important results obtained by them. However, papers \cite{Quarles-03,Brazil} did not investigate the ultra-soft x-ray photon energy region.

The presented above analysis of the literature shows that the study of BS for atomic targets, is not only interesting by itself, but is useful for studies of BS generated in electron scattering on solid-state targets (except for the above references, see e.g. \cite{13,14}). In addition, these data are useful in studies of BS in the scattering of electrons by molecules. As an example serves Ref.~\cite{India}, that presents investigation of BS generated in electron scattering upon molecules CH$_4$.

Since, as already noted, the greatest discrepancy between the experimental data presented in \cite{Quarles-03, Brazil} and the theory is exists in the small photon energies region, the data obtained in Refs.~\cite{Nechay0, Nechay1} are of particular interest. These works present results of absolute, not relative, as in \cite{Peterson}, measurements of BS cross sections values for ultra-soft x-ray photons ($\lambda = 7..18$~nm) generated in scattering of electrons with energies of 0.4..2~keV on Ar, Kr, and Xe atoms. The experimental data differ by a factor 3..4 from the calculations of Pratt \textit{et al.} \cite{Pratt77, Pratt81}, carried out with an accurate account of the phases of the electron wave functions in the static atomic potential, but with the radial dependence of these functions that corresponds to the free motion of the scattered electron. Note, however, that the papers \cite{Nechay0, Nechay1} present BS data for electrons of intermediate energies, where the Born approximation is invalid. Let us remind that the first Born approximation is sufficient only if the speed of incoming electron is much higher than the average speed of all atomic electrons, including the innermost ones. Corresponding numerical examples are given, \emph{e.~g.}, in Ref. \cite{JPB-08}.

The important results reported in Refs. \cite{Nechay0,Nechay1}, have no theoretical explanation yet. In this paper, some of these results, namely, the values of BS cross sections, including the limiting extrapolated values at infinitely large wavelengths of BS photons, 
we interpret on the basis of the soft-photon approximation (SPA) of quantum electrodynamics \cite{AB},
\begin{eqnarray}\label{AB1}
\omega\ll E/\hbar,
\end{eqnarray}
where $\omega$ is the frequency of the BS photon and $E$ is the energy of the electron in the initial state. Previously, the SPA was used to interpret ultra-soft x-ray spectra in Refs.~\cite{PRL-05,Verkh-06,Verkh-90,Zon-95}. Recall that the SPA takes into account both elastic and inelastic scattering of an electron by an atom, as well as polarization BS. To be able to take into account the polarization BS, it is necessary that the energy of the BS photon be much less than the binding energies of atomic electrons that  lie off shall of the mass surface, unlike free electrons \cite{AB}. However, in the case considered here, this condition is not fulfilled; therefore, the polarization BS is not considered here by the SPA.

\section{Basic formula}

In the frame of SPA and in non-relativistic approximation, the double differential cross section of BS
for electron scattering on a certain target is presented as the product of the differential cross section of electron scattering on the same target without radiation, $d\sigma_S$, and the probability of photon radiation $dw_\gamma$ \cite{AB},
\begin{eqnarray}\label{AB4}
&&d\sigma=d\sigma_S dw_\gamma, \\\nonumber && dw_\gamma =\frac{\alpha}{4\pi^2}\left[\frac{\textbf k}{\omega}(\textbf v -\textbf v') \right]^2 \frac{d\omega}{\omega}\,d\Omega_\gamma.
\end{eqnarray}
where $\alpha\simeq 1/137$ is the fine structure constant, $\textbf v$ and $\textbf v'$ are electron velocities
in the initial and final states, respectively, $\textbf k$ is photon momentum, and $d\Omega_\gamma$ is the solid angle of photon departure. We employ the relativistic system of units in which $\hbar=c=1$. In equation \eqref{AB4} averaging over the electron spins in the initial state,
summation over its spins in the final state and summation over the photon polarizations are carried out.
If the direction of flight of the electrons is not fixed as in the experiment \cite{Nechay0,Nechay1}, then it is necessary to perform integration over this direction in the equation \eqref{AB4} . To do this, we direct the axis $z$ along
the vector $\textbf v$ and write $d\sigma_S$ as
\begin{eqnarray}\label{AB5}
d\sigma_S(\theta)=\sigma_S(\theta) d\Omega_{\textbf v'},
\end{eqnarray}
where $\theta$ is the polar angle of the vector $\textbf v'$. We assume a spherical symmetry of the target, so
$d\sigma_S$ does not depend on the azimuthal angle of $\textbf v'$. Substituting \eqref{AB5} in \eqref{AB4}, we obtain,
\begin{eqnarray}\nonumber
d\sigma=\frac{\alpha}{4\pi^2\omega}\sigma_S(\theta) \left[v^2\cos^2\Theta - \frac{2v}{\omega} (\textbf{kv}') \cos\Theta \right.\\\nonumber  \left.+\frac{(\textbf{kv}')^2}{\omega^2} \right] d\Omega_{\textbf v'} d\omega d\Omega_\gamma 
= \frac{\alpha}{4\pi^2\omega}\sigma_S(\theta) \left[v^2\cos^2\Theta -  \right.\\\nonumber
-2vv'\cos\Theta (\sin\Theta\sin\theta\cos\varphi+\cos\Theta\cos\theta)+ \\\nonumber \left.+ v'^2(\sin\Theta\sin\theta\cos\varphi+\cos\Theta\cos\theta)^2\right]d\Omega_{\textbf v'} d\omega d\Omega_\gamma.
\end{eqnarray}
where $\Theta$ is the polar angle of the vector $\textbf k$ (the angle between the vectors $\textbf v$ and $\textbf k$),
the azimuthal angle of the vector $\textbf k$ is assumed to be zero, $\varphi$ is the azimuthal angle of the vector
$\textbf v'$. In experiments \cite{Nechay0, Nechay1} the angle $\Theta\simeq 83^\circ$. After integration by angle
$\varphi$ we obtain,
\begin{eqnarray}\label{AB6}
&&d\sigma= \frac{\alpha}{2\pi\omega}\sigma_S(\theta) \left[v^2\cos^2\Theta
-2vv'\cos^2\Theta \cos\theta+ \right.\\\nonumber &&\left.+ v'^2 \left(\frac12\sin^2\Theta\sin^2\theta+\cos^2\Theta\cos^2\theta \right) \right] \sin\theta d\theta d\omega d\Omega_\gamma.
\end{eqnarray}

Given the relation \eqref{AB1}, we can consider $\textbf v'\simeq\textbf v$, which allows us to write the equation
\eqref{AB6} in the following form,
\begin{eqnarray}\label{AB7}
&&d\sigma= \frac{\alpha v^2}{2\pi\omega}\sigma_S(\theta)f(\theta) \sin\theta d\theta d\omega d\Omega_\gamma,\\\nonumber &&f(\theta)= \cos^2\Theta (1 -\cos\theta)^2 + \frac12\sin^2\Theta\sin^2\theta.
\end{eqnarray}

The BS cross section \eqref{AB7} determines the radiation intensity $I$ in the solid angle $d\Omega_\gamma$ and in the wavelength interval $d\lambda\; (\lambda=2\pi/\omega)$ by the following equation,
\begin{eqnarray}\label{AB8}
\frac{dI}{d\lambda d\Omega_\gamma}= \frac{\alpha v^3 n_e}{\lambda^2} (n_a\Delta V) \int_0^\pi \sigma_S(\theta) f(\theta) \sin\theta d\theta,
\end{eqnarray}
where $n_a, n_e$ are densities of atoms and electrons, respectively, $\Delta V$ is effective volume of
interaction of electrons with atoms, from which radiation originate. To obtain $dI/d\lambda\,d\Omega_\gamma$ in normal units, one has to multiply Eq. \eqref{AB8} by $\hbar/c$.

Note that in Ref.~\cite{Peterson} in the region of large wavelengths, the dependence $dI\sim 1/\lambda^\alpha$ with $1.8<\alpha<2.7$ was observed. This dependence does not contradict dependence $dI\sim 1/\lambda^2$, which follows from Eq.~\eqref{AB8}.

\section{Results and discussion}

\begin{figure}
\includegraphics[scale=0.475]{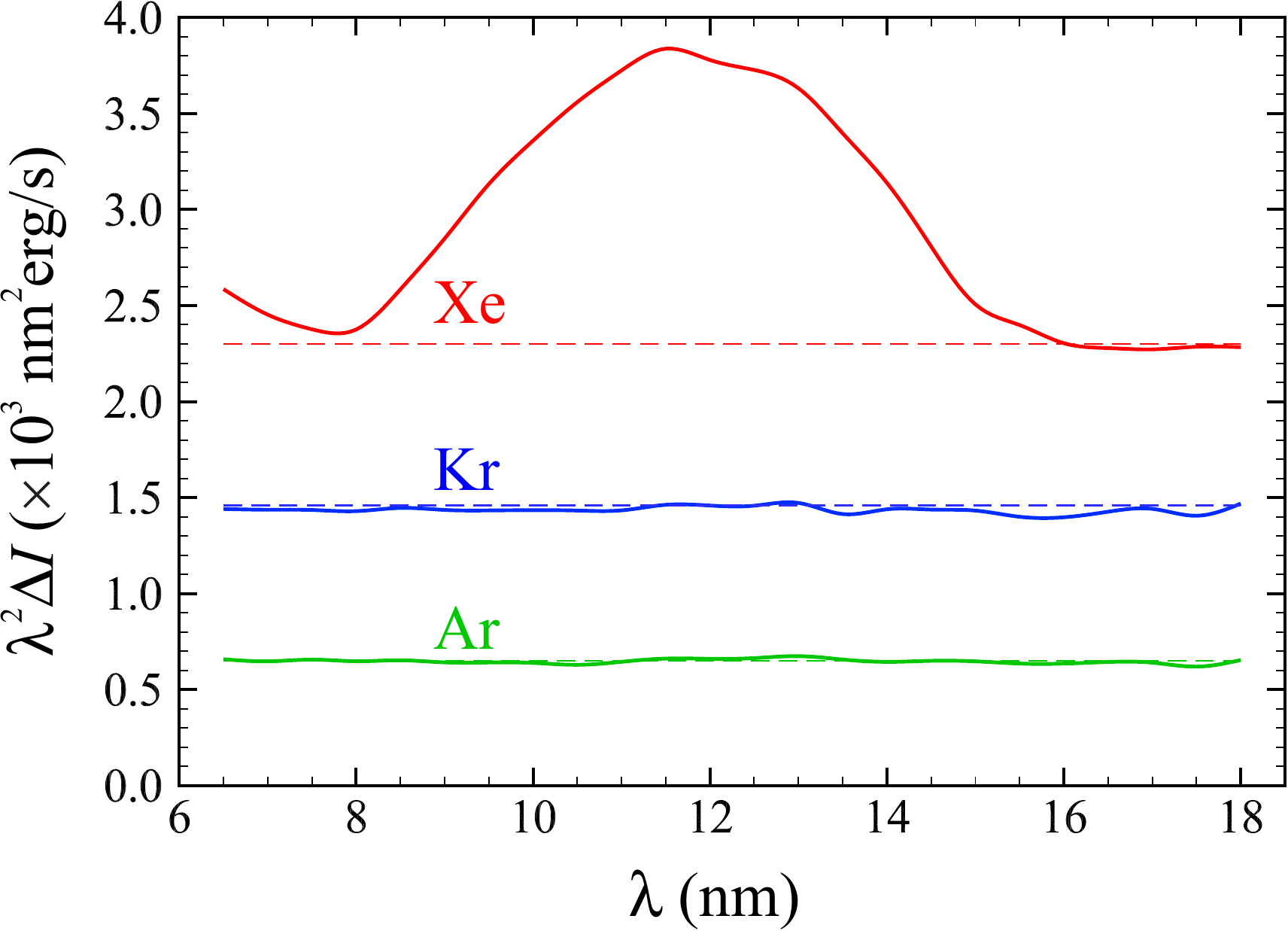}
\caption{\label{Fig1}
(Color online) The dependence of $\lambda^2\Delta I$ upon the radiation wavelength $\lambda$. We took the plots from Fig.~4 of Ref. \cite{Nechay1}, which presents data for electrons with the initial energy 600~eV. Horizontal dashed lines correspond to constant values of $\lambda^2\Delta I$. Non-monotonic dependence of plot for Xe is connected to the polarization BS \cite{Tsyt,JPB1990}.}
\end{figure}

In experiments \cite{Nechay0, Nechay1}, the volume of interaction of electrons with atoms $\Delta V\simeq 0.1$~cm$^3$, and radiation was collected from the finite solid angle $\Delta\Omega_\gamma\simeq 1.7\times10^{-3}$~sr in the wavelength interval $\Delta\lambda\simeq 0.1$~nm. The intensity of BS emitted from this spatial region and in the specified wavelength interval is denoted by $\Delta I$. It follows from Eq. \eqref{AB8} that for large wavelengths, the value of $\lambda^2 dI/d\lambda\,d\Omega_\gamma$ is independent upon $\lambda$. We see it from Fig.~\ref{Fig1} that depicts the values of $\lambda^2 \Delta I$ obtained from Fig.~4 in Ref.~\cite{Nechay1}. The impressive maximum in BS cross section dependence on the radiation wavelength observed in the case of Xe is a direct manifestation of the polarization BS. It is possible that the relatively small growth of $\lambda^2\Delta I$ at small $\lambda$  in Fig.~\ref{Fig1} has the same origin. Note, however, that such non-monotonic behavior is located, as it should be for polarization BS, only in a limited wavelength range, where polarizability of Xe is high (giant resonance in photoabsorbtion \cite{Tsyt,JPB1990}. One should have in mind that polarizabilities of Ar and Kr in the considered $\lambda$  region are on the contrary small. As one can see from Fig.~\ref{Fig1}, the SPA is applicable for Xe in the wavelength range 16..18~nm, where $\lambda^2 dI/d\lambda\,d\Omega_\gamma$ is almost constant.

To calculate $\sigma_S (\theta)$ we limited ourselves to elastic scattering and we used the data from Refs. \cite{Jab-04,Jab-10}. Note that the elastic scattering cross section $\sigma_S (\theta)$ has a pronounced maximum at small angles, where the function $f(\theta)$, defined by Eq. \eqref{AB7}, has a minimum at the value $\Theta$ used in the experiment. In the maximum of the function $f(\theta)$, which is at $\theta\simeq 100^\circ$, the cross section $\sigma_S (\theta)$ is less than its maximum by 2..3 orders of magnitude. Since the functions $\sigma_S(\theta)$ and $f(\theta)$ enter Eq.~\eqref{AB8} as a product under the integral over $\theta$, the accuracy of theoretical calculations of the cross section should be sufficiently high. Previously, this circumstance was noted in Ref. \cite{PRL-05}.

The results of theoretical calculations are given in Table~\ref{Table1}. The same Table includes experimental data from Fig.~\ref{Fig1}. One should have in mind that the experimental error of absolute measurements is 40\%.
As one can see, for Kr and Xe, the theoretical values lie within the range of experimental values and slightly out of it for Ar.

\begin{table}
\caption{\label{Table1} Experimental and theoretical values of $\lambda^2 \Delta I$, which does not depend on $\lambda$,~nm$^2$erg/s.}
\begin{ruledtabular}
\begin{tabular}{ccc}
Atom & Experiment & Theory \\
\hline
Ar  & $650\pm 260$ & 921   \\
Kr  & $1460\pm 584$ & 1158  \\
Xe  & $2300\pm 920$ & 1725  \\
\end{tabular}
\end{ruledtabular}
\end{table}

\section{Conclusion}

The presented results demonstrate for the first time the quantitative agreement of theoretical and experimental ordinary BS cross sections in electron-atomic collisions at low energies of BS photons. This result does not apply to the polarization BS, for the manifestation of which requires the fulfillment of special conditions. These conditions are well-known \cite{Tsyt, Astapenko, KS-book} and are (a) high optical polarizability of atoms (optical breakthrough of alkaline metals was the first experimental proof of polarization BS \cite{Zon-77}), or high X-ray polarizability as in Xe atom (Fig.~\ref{Fig1}), (b) laser-assisted electron scattering on very small angles \cite{Yam,28,29} etc.

In addition, these same results present an independent verification of high accuracy of differential cross sections of elastic scattering of electrons on atoms that we take from Refs. \cite{Jab-04, Jab-10}.

\begin{acknowledgments}
The authors express their deep gratitude to A.A.Tkachenko for the detailed discussion of the experimental results presented in the Refs.~\cite{Nechay0, Nechay1}. A.~S.~K. and B.~A.~Z. are also grateful Russian Ministry of higher education and science for financial support (State Assignment No.~3.1761.2017/4.6).
\end{acknowledgments}

\end{document}